\documentclass{kapproc}
\usepackage{graphicx}

\def\laeq{\stackrel{<}{\scriptstyle \sim}}

\begin{document}

\articletitle{The Galactic Kinematics of Mira Variables}

\author{Michael Feast}
\affil{Astronomy Department, University of Cape Town, Rondebosch,7701,\\ 
South Africa}
\email{mwf@artemisia.ast.uct.ac.za}

\begin{abstract}
The galactic kinematics of Mira variables derived from radial velocities,
Hipparcos proper motions and an infrared period-luminosity relation are
reviewed. Local Miras in the 145-200day period
range show a large asymmetric drift and a high net outward
motion in the Galaxy. Interpretations of this phenomenon are considered and
(following Feast and Whitelock 2000) it is suggested that they are outlying
members of the bulge-bar population and indicate that this bar extends beyond
the solar circle. 
\end{abstract}


\section{Introduction}
  The galactic kinematics of Mira variables have for a long while been
of importance in helping us understand both the nature and evolution of
Miras as well as the structure of our own Galaxy. This paper is primarily
concerned with the second point - what do we learn from Miras about
galactic structure? The paper also concentrates on local kinematics,
leaving out a detailed discussion of the kinematics of the galactic
bulge, much work on which has of course been done here in Japan. It
has been possible to take a fresh look at the local kinematics of
Miras using Hipparcos astrometry, extensive new infrared photometry
and published radial velocities. This was done in a series of three papers
(Whitelock, Marang and Feast (2000), paper I: Whitelock and Feast (2000),
paper II: Feast and Whitelock (2000a), paper III). The present paper
summarizes some of the relevant results from these papers and extends
the discussion of the kinematics.
 
\section{The Period-Kinematic Sequence}
 It has been known for many years that there is a relation between
the kinematics and the periods of Mira variables in the general solar
neighbourhood. In particular the asymmetric drift and the velocity
dispersion increase as one goes from longer to shorter periods.
There was however an anomaly.  The shortest period group (periods
less than $\sim 150$ days) differed strongly from the general trend
(Feast 1963). Following an initial study by Hron (1991), it was shown in
paper I (see also Whitelock 2002) that combining infrared and Hipparcos
magnitudes one sees two sequences of Miras at shorter periods 
in the
period - colour plane, 
the SP-red and SP-blue stars,
and this must be taken into account when discussing
the kinematics.

\begin{figure}[t]
\begin{center}
\includegraphics[height=10cm]{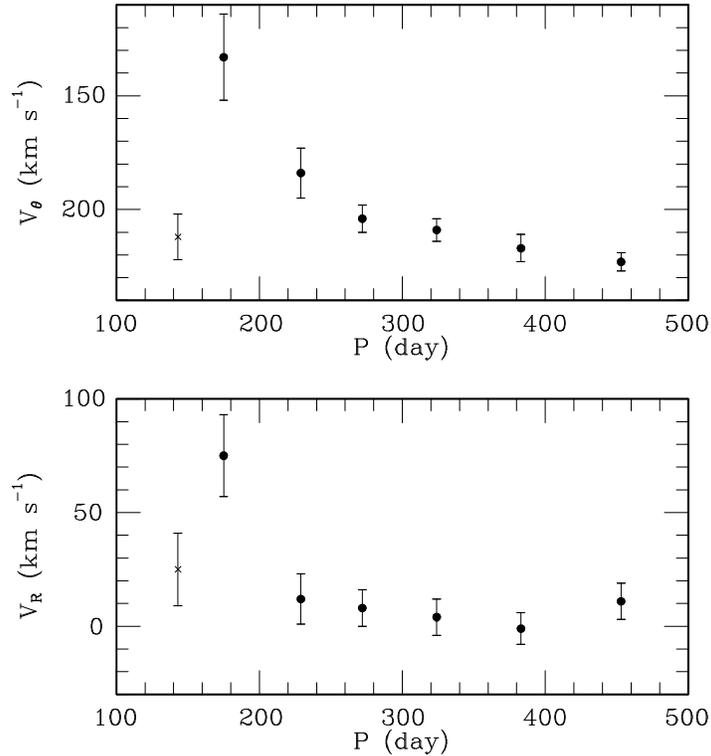}
\caption{(a) The mean $V_{\theta}$ in various period groups plotted against
the mean period of the group.
(b) The mean $V_{R}$ in various period groups plotted against the mean
period of the group. The cross represents the SP-red variables in both
figures.}
\end{center}
\end{figure} 
Figure 1a shows a plot of mean velocity of Miras in the direction of
galactic rotation, $V_{\theta}$, in period groups. 
This plot (data in paper III) is a distinct improvement on earlier work.
It depends on space motions rather than just radial velocities and the
SP-red stars have been grouped separately and are denoted by a cross.
The figure shows that 
at the longest periods the stars of the main Mira sequence are moving at
close to the circular velocity of galactic rotation, which is taken as
231 $\rm km.s^{-1}$ (Feast and Whitelock 1997). There is a clear dependence
of mean rotational velocity on period for this main Mira sequence with 
the short period (145-200 day) Miras having a large asymmetric drift 
($-97 \pm 20 \rm km.s^{-1}$).\\ 
The main Mira sequence now seems reasonably well
understood. The infrared colour-period relation for this sequence is the same
as that for Miras in globular clusters (paper I and Whitelock 2002), strongly
suggesting a similarity between the field and cluster Miras 
over the period range in common. Not only do the clusters show that
their Miras lie at the tip of the AGB, they also show that there is a
period-metallicity relation (e.g. Feast and Whitelock 2000b). This allows us to
study the galactic kinematics of old populations as a function of metallicity
(and possibly also age) over a range of ages and metallicities where there
are few if any useful, precise, tracers. 
For instance globular clusters containing Miras are often classed as ``disc"
clusters and treated together in kinematic discussions. However the field
Miras show that there is a considerable variation of kinematic properties
in the period/metallicity range of relevance to these clusters.\\
Where do the SP-red stars fit in? A full answer to this question may come when
more members of the class are identified and when good individual
parallaxes are available for a significant number. The best guess at present
(see e.g. Whitelock 2002) is that, unlike the stars of the main Mira sequence,
they are not at the end of their AGB lives but that they will evolve into 
longer period Miras of the main Mira sequence.\\
\section{Local Miras and a Galactic Bar}
A particularly interesting result in paper III is the evidence of a
net radial, outward motion ($V_{R}$) for Miras in the solar neighbourhood. This
is shown as a function of period in figure 1b
There is evidence of a small outward motion
over most of the period range of the Miras. However it is
very marked in the group with periods in the range 145 to 200 days where one
finds, $V_{R} = +75 \pm 18 \rm km.s^{-1}$ for a mean period of 173 days. 
Note that in view of the discussion in section 2, the SP-red Miras are
omitted in the discussion of the present section.\\
There has been evidence in the literature for many years that some groups
of old stars in the solar neighbourhood 
show a predominant, but modest, outward motion in the Galaxy, though this
has not always been recognized. The Hipparcos results, giving parallaxes
and proper motions for many thousands of common stars have enabled this
to be studied in greater detail. It is now clear that there is a
group of old stars with a small asymmetric drift and a modest outward
velocity $(V_{\theta} = \sim 190 \rm km.s^{-1}$, 
$V_{R} = \sim +30 \rm km.s^{-1}$ (Dehnen 1998,
1999, 2000, Fux 2001). Fux has referred to these stars
as forming the "Hercules Stream". Dehnen, Fux and also Quillen (2002) each 
suggest that the stars of the Hercules Stream were originally in
circular orbits which have been perturbed into their present state by
the influence of a (pre-existing) galactic bar. However, the detailed
dynamic processes envisaged differ from author to author.\\ 
 Fux has
suggested that the short period Mira group just discussed belongs to the
Hercules Stream. However whilst the Hercules Stream consists of old
stars with a wide range of metallicities (Raboud et al. 1998)
the large local outward motion  in the case of Miras is confined to a restricted
range in period, and therefore also in metallicity. The effect is
also more extreme than in the Hercules group. This is shown in Figure 2
which plots $V_{\theta}$ against $V_{R}$ for individual Miras in the
145 to 200 day group. 

\begin{figure}[h]
\begin{center}
\includegraphics[height=7cm]{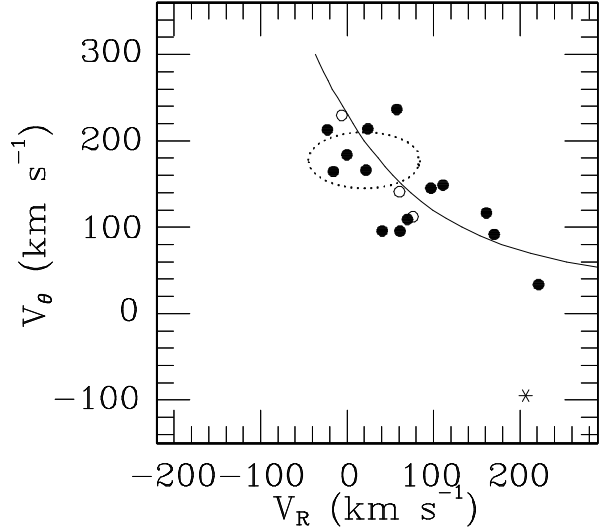}
\caption{A plot of $V_{\theta}$ against $V_{R}$ for local Miras with periods
in the range 145 to 200 days. The open circles denote stars for which the
standard error of a velocity component is greater than $20 \rm km.s^{-1}$.
The asterisk show the position of S Car which is on a highly eccentric
retrograde orbit. The curve and dotted oval are discussed in the text.}
\end{center}
\end{figure}
The dotted oval shows the extent of the Hercules
Stream as delimited by Dehnen (2000, fig 9). Evidently these Miras do 
not concentrate in the region of the Hercules Stream. \\
It is clear that some of
the stars in figure 2 are on highly eccentric orbits. About half the
sample have perigalactic distances less than about 2kpc and some at least will
pass through the galactic bulge. Despite the small number of stars in this
diagram, there is also evidence of a relation between $V_{\theta}$ and
$V_{R}$. In paper III we derived the curve shown in the fig 3 which makes
the first order assumption that the stars are moving in simple
elliptical orbits of different eccentricities but with their major axes
aligned. There is then only one free parameter; the position of
this major axis with respect to the sun-centre line. The curve is drawn for the
best fitting value of this position angle ($ 17^{+11}_{-4}$ degrees). It 
provides a rather good fit to the data and since this angle is, within
the uncertainties, similar to that proposed for the bar in the galactic
bulge, we suggested that these 145-200 day Miras in the general
solar neighbourhood were outlying members  of the bar population itself. 
If this is the
case the bar population extends out to beyond the solar circle.\\
This discussion depends on the Miras in the general solar neighbourhood
and is thus limited in number. In a recent paper Kharchenko et al. (2002)
have suggested that over a larger volume of space, the mean $V_{R}$ for
145-200 day Miras is near zero. More details of their work are need for
a proper discussion. However the following points are worth noting. (1)
the distances adopted by Kharchenko et al. 
are obtained from visual magnitudes and the reddenings from 
a model. They must thus be rather uncertain. (2) A
certain amount of trimming is carried out (velocities further from 
means than $3\sigma$ are rejected). (3) Complications in the analysis
will arise when one goes to a large volume if the Galaxy is not
axi-symmetrical. (4) No distinction is made in Kharchenko et al.
between the SP-red and SP-blue stars discussed above and this is particularly
important in the period range under discussion.\\
Whilst therefore the large scale picture remains unclear, it seems rather
remarkable that the
nearby Miras in the period range 145-200 days with radial velocities,
Hipparcos proper motions and infrared photometry show a marked asymmetry in
$V_{R}$. As figure 2 shows, all the stars in this group with
($V_{\theta} \laeq \rm 160 km.s^{-1}$) have positive values of $V_{R}$.
For an axi-symmetrical galaxy there should be a symmetrical distribution
of $V_{R}$ about zero in this figure at any given $V_{\theta}$. 
The deviation from such a
distribution is sufficiently striking that it seems difficult to
attribute it entirely to chance.
\begin{acknowledgments}
I would like to thank Professor Nakada (University of Tokyo) and
the organizers of this meeting for making my attendance possible.
This paper depends on work done in collaboration with Patricia Whitelock.
\end{acknowledgments}

\section{Discussion}
\noindent{\bf Habing}\\
   In your calculations of the galactic orbit, did you assume the bar is
stationary? It may rotate.\\
{\bf Feast}\\
 For a rotating bar one needs to assume the simple elliptical
orbits precess. One is then concerned with the present orientations
of the Mira orbits and the bar. So the conclusions are not affected\\
{\bf van Langevelde}\\
1. You find a best fit of $\phi$, the angle between the major axis of the 
Mira orbits and the bar, but there must be a whole range of eccentricities.
What is the range of perigalactic distances?\\
2. What makes the 145-200 day group special? Would that indicate something
about the population/age of the bar?\\
{\bf Feast}\\
1. The distribution in figure 2 is essentially a distribution in
eccentricity. About half the stars in that plot go within 2 kpc of the
centre. The exact perigalactic distance depends of course on the mass
model.\\
2. That is not entirely clear. There are Miras with a range of periods in
the bulge. At periods longer than about 200 days, the
local population is dominated by variables on much more nearly
circular orbits. Possibly this is because the galactic density gradient of
Miras is a function of period.\\
{\bf Nakada}\\
Do the bulge Miras dominate the short period (P 145-200 days) 
group in the solar
neighbourhood?\\
{\bf Feast}\\
About half the Miras in this group 
(SP-red stars omitted) have perigalactic distances sufficiently
small that we can probably say they belong to a bulge population. However
there is no evidence at present that the stars in this period group which are
on more nearly circular orbits are different physically from 
those on highly eccentric orbits.
In that sense they can perhaps all be regarded as a bulge type population.\\
\end{document}